\documentclass[useAMS,usenatbib,usegraphicx,letterpaper]{mymn2e}
                                   
\newcommand{\be}{\begin{equation}}
\newcommand{\ee}{\end{equation}}
\newcommand{\ba}{\begin{eqnarray}}
\newcommand{\ea}{\end{eqnarray}}

\newcommand{\etal}{et al.}

\newcommand{\ie}{i.e.}
\def\simless{\mathbin{\lower 3pt\hbox
  {$\rlap{\raise 4pt\hbox{$\char'074$}}\mathchar"7218$}}}
\def\simgreat{\mathbin{\lower 3pt\hbox
  {$\rlap{\raise 4pt\hbox{$\char'076$}}\mathchar"7218$}}}

\newcommand\yj{\mbox{$Y\!-\!J$}}%   % Y-J
\newcommand\jh{\mbox{$J\!-\!H$}}%   % J-H
\newcommand\jhk{\mbox{$J\!H\!K$}}%   % JHK
\newcommand\yjhk{\mbox{$Y\!J\!H\!K$}}%   % 
\newcommand\zyjhk{\mbox{$Z\!Y\!J\!H\!K$}}%   % 
\newcommand{\object}{ULAS~J0034$-$00}
\newcommand{\mobject}{2MASS~J0415$-$09}
\newcommand{\gobject}{Gl~570D}
\newcommand{\tf}{\mbox{$T_{\rm{eff}}$}}

\title
[A very cool brown dwarf]
{A very cool brown dwarf in UKIDSS DR1}
\author
[S. J. Warren, \etal]
{S. J. Warren$^1$\thanks{E-mail: s.j.warren@imperial.ac.uk},
D. J. Mortlock$^1$,
S. K. Leggett$^2$, 
D. J. Pinfield$^3$,   
D. Homeier$^4$, \newauthor
S. Dye$^5$,
R. F. Jameson$^6$, 
N. Lodieu$^7$,   
P. W. Lucas$^3$, 
A. J. Adamson$^8$,
F. Allard$^{9,10}$, \newauthor
D. Barrado y Navascu\'{e}s$^{11}$,
M. Casali$^{12}$,
K. Chiu$^{13}$,
N. C. Hambly$^{14}$, \newauthor 
P. C. Hewett$^{15}$,
P. Hirst$^2$,
M. J. Irwin$^{15}$,
A. Lawrence$^{14}$,
M. C. Liu$^{16}$\thanks{Alfred P. Sloan Research Fellow}, \newauthor
E. L. Mart\'{i}n$^7$,
R. L. Smart$^{17}$,
L. Valdivielso$^7$,
B. P. Venemans$^{15}$
\vspace{7mm}\\
$^1$Astrophysics Group, Imperial College London, Blackett Laboratory,
  Prince Consort Road, London, SW7 2AZ, U.K.\\
$^2$Gemini Observatory, Northern Operations Center, 670 North A'ohoku
  Place, Hilo, HI96720, U.S.A. \\ 
$^3$Centre for Astrophysics Research, Science and Technology Research 
Institute, University of Hertfordshire, Hatfield, AL10 9AB, U.K. \\
$^4$Institut f\"{u}r Astrophysik, Georg-August-Universit{\"a}t, 
  Friedrich-Hund-Platz 1, 37077 G{\"o}ttingen, Germany\\ 
$^5$Cardiff School of Physics and Astronomy, Cardiff University,
Queens Buildings, The Parade, Cardiff, CF24 3AA, U.K. \\
$^6$Department of Physics and Astronomy, University of Leicester,
Leicester, LE1 7RH, U.K. \\
$^7$Instituto de Astrof\'{i}sica de Canarias, V\'{i}a L\'{a}ctea s/n, E-38205
La Laguna, Tenerife, Spain \\
$^8$Joint Astronomy Centre, 660 North A'ohoku Place,
Hilo, HI96720, U.S.A.\\
$^9$Centre de Recherche Astrophysique de Lyon, UMR5574, CNRS,
Universit\'{e} de Lyon, \'{E}cole Normale Sup\'{e}rieure, \\
~~46 All\'{e}e d'Italie, F-69364 Lyon Cedex 07, France \\
$^{10}$Institut d'Astrophysique de Paris, UMR7095, CNRS, Universit\'{e}
Pierre et Marie Curie, 98$^{bis}$ Boulevard Arago, F-75014 Paris, France \\
$^{11}$Laboratorio de Astrof\'{i}sica Espacial y F\'{i}sica Fundamental
(LAEFF-INTA), P.O. Box 50727, E-28080 Madrid, Spain \\
$^{12}$ESO, Karl-Schwarzschild-Str. 2, D-85748 Garching bei M\"{u}nchen, 
Germany \\
$^{13}$Astrophysics Group, School of Physics, University of Exeter,
Stocker Road, Exeter, EX4 4QL, U.K. \\
$^{14}$Scottish Universities Physics Alliance (SUPA),
Institute for Astronomy, School of Physics, University of Edinburgh, \\
~~Royal Observatory, Blackford Hill, Edinburgh, EH9 3HJ, U.K. \\
$^{15}$Institute of Astronomy, Madingley Rd., Cambridge, CB3 0HA, U.K.\\
$^{16}$Institute for Astronomy, University of Hawai'i, 2680 Woodlawn Drive,
Honolulu, HI96822, U.S.A. \\
$^{17}$Osservatorio Astronomico di Torino, 10025 Pino Torinese, Italy \\
}

\begin{document}

\date{}

\pagerange{\pageref{firstpage}--\pageref{lastpage}} \pubyear{2007}

\maketitle

\label{firstpage}

\begin{abstract}
We report the discovery of a very cool brown dwarf, ULAS
J003402.77$-$005206.7 (\object), identified in the UKIRT Infrared Deep
Sky Survey First Data Release. We provide optical, near-infrared, and
mid-infrared photometry of the source, and two near-infrared
spectra. Comparing the spectral energy distribution of \object\ to
that of the T8 brown dwarf 2MASS J04151954$-$0935066 (\mobject), the
latest-type and coolest well-studied brown dwarf to date, with
effective temperature $ \tf\sim750$\,K, we find evidence that
\object\ is significantly cooler.  First, the measured values of the
near-infrared absorption spectral indices imply a later
classification, of T8.5. Second, the $H\!-$[4.49] colour provides an
empirical estimate of the effective temperature of $540<\tf<660$\,K
($\pm2\sigma$ range). Third, the $J$- and $H$-band peaks are somewhat
narrower in \object, and detailed comparison against spectral models
calibrated to \mobject\ yields an estimated temperature lower by
$60<\Delta\tf<120$\,K relative to \mobject\, i.e. $630<\tf<690$\,K
($\pm2\sigma$), and lower gravity or higher metallicity according to
the degenerate combination $-0.5<\Delta($log\,$g-2$[m/H]$)<-0.25$
($\pm2\sigma$). Combining these estimates, and considering
systematics, it is likely the temperature lies in the range
$600<\tf<700$\,K. Measurement of the parallax will allow an
additional check of the inferred low temperature.  Despite the low
inferred \tf\ we find no evidence for strong absorption by NH$_3$ over
the wavelength range $1.51$--$1.56$\,$\mu$m.  Evolutionary models
imply that the mass and age are in the ranges 15--36 $M_{\rm Jup}$ and
0.5--8\,Gyr, respectively. The measured proper motion, of
$(0\farcs37\pm0\farcs07)$/yr, combined with the photometrically
estimated distance of 14--22~pc, implies a tangential velocity of
$\sim$30~km\,s$^{-1}$, a value consistent with expectation for the
inferred age.  \object\ is significantly bluer than \mobject\ in \yj,
so future searches should allow for the possibility that cooler T
dwarfs are bluer still.

\end{abstract}

\begin{keywords}
brown dwarfs -- surveys
\end{keywords}

%%%%%%%%%%%%%%%%%%%%%%%%%%%%%%%%%%%%%%%%%%%%%%%%%%%%%%%%%%%%%%%%%%%%%%%%%%%%%%

\mbox{}
\vspace{2 cm}

\section{Introduction}

In the decade since the first discovery of a T dwarf
\citep[Gl\,229,][]{nak95} over one hundred have been identified.  Most
have been discovered in the field, in large surveys, especially the
Sloan Digital Sky Survey \citep[SDSS;][]{str99,yo00} and the Two
Micron All Sky Survey \citep[2MASS;][]{bur99,sk06}. \citet{bur06a}
provide a classification scheme that uses comparison against a set of
spectral templates over the range T0 to T8, and measurement of a set of
spectral indices that quantify the depth of H$_2$O and CH$_4$
absorption features in the $J$, $H$ and $K$ bands. The known T dwarfs
have temperatures in the range $700 \, \simless T_{\rm{eff}} \simless
1500 \, {\rm{K}}$ \citep{golimowski04}. To the end of 2006 only six
dwarfs classified T7.5 or T8 had appeared in the literature.
\citet{sau06} and \citet{sau07} made a detailed study of three of
these cool T dwarfs, combining near- and mid-infrared photometry and
spectroscopy to infer effective temperatures of $725 \, {\rm{K}} \leq
T_{\rm eff} \leq 950 \, {\rm{K}}$, surface gravities of $4.8 \leq
\log[g / (1 \, {\rm{cm}} \, {\rm{s}}^{-2})] \leq 5.4$, and masses of
$25 \, M_{\rm{Jup}} \leq M \leq 65 \, M_{\rm{Jup}}$.

The temperature range $T_{\rm eff} \simless 700 \, {\rm{K}}$ remains
unexplored at present, but models such as those of \citet{burrows03}
can be used to design searches for brown dwarfs beyond T8. Such very
cool brown dwarfs are expected to have very low luminosities, to be
extremely red in $z-J$, and, at least for $T_{\rm eff} \simgreat 400
{\rm{K}}$, to be blue in $J-K$. Therefore they will be faint in both
the $z$ and $K$ bands. Shortward of $1\,\mu$m, optical searches such
as SDSS are clearly unsuited for detecting such very cool brown
dwarfs. However, a {\em JHK} survey, such as 2MASS, is also far from
ideal, as cool T dwarfs will typically be undetected in {\em
  K}. Samples of candidate cool T dwarfs selected as blue in \jh, but
detected in only these two bands can suffer from a high degree of
contamination.  As described by \citet{he06}, the $Y$ band
\citep{hill,wa02}, between $z$ and $J$, is a valuable addition to the
$J$ and $H$ bands in a search for very cool T dwarfs. Fig.~2 of
\citet{he06}, and Fig.~1 of \citet{lodieu}, illustrate the fact that T
dwarfs in the spectral range T3--T8 are much redder in \yj\ than main
sequence stars of similar \jh\ colour, meaning that a broader range of
spectral type of T dwarf may be identified than in a simple
\jh\ search, and with greatly reduced contamination. Whilst there is
some uncertainty about the near-infrared colours of T8+ dwarfs, they
should lie in a sparsely populated region of the \yj\, vs. \jh\,
two-colour diagram, and so the extension of the T dwarf sequence to
later types should be readily identifiable using these colours.

A wide-field near-infrared survey that is substantially more sensitive
than 2MASS is needed to find such faint red objects, and this was one
of several considerations which led to the development of the UKIRT
Infrared Deep Sky Survey \citep[UKIDSS;][]{lawrence07}.  UKIDSS is a
set of five surveys using the UKIRT Wide Field Camera
\citep[WFCAM;][]{casali07} to obtain imaging of $\sim$20 per cent of
the sky in a number of bands selected from the $Z\!Y\!J\!H\!K$ set --
see \citet{lawrence07} for details. UKIDSS began in 2005, and is
estimated to take seven years to complete. Three data releases have
now taken place: the Early Data Release \citep[EDR;][]{dye06}; the
First Data Release \citep[DR1;][]{wa07a}; and the Second Data Release
\citep[DR2;][]{wa07b}.

Of the five UKIDSS surveys, The Large Area Survey (LAS) is best suited
to searching for very cool brown dwarfs.  The LAS plans to cover
$\sim$4000$\, {\rm{deg}}^2$ within the SDSS footprint in the {\yjhk}
bands and has an average depth (to a signal--to--noise ratio (S/N) of
five for point--sources) of $K \simeq18.2$.  Critically, this is $2.7$
magnitudes deeper than 2MASS, meaning that the $190 \, {\rm{deg}}^2$
covered in {\yjhk} by the LAS DR1 fields already probes a volume one
fifth that of 2MASS.\footnote{The gain in depth in the {\em H} band,
  which sets the volume in a search for very cool brown dwarfs, is
  very similar to the gain in {\em K}.}

A search for red point--sources in the UKIDSS EDR and DR1 LAS data has
already yielded nine new spectroscopically--confirmed T dwarfs
\citep{kendall,lodieu} over the range T4 to T7.5. Here we report the
discovery of a very cool brown dwarf, ULAS J003402.77$-$005206.7
(hereafter \object), that appears to have a later type, lower
effective temperature and possibly lower mass than any other known T
dwarf. The discovery process and follow-up photometric, astrometric,
and spectroscopic observations are described in Section~\ref{obs}.  In
Section~\ref{spectype} we use a $1-2.5$\,$\mu$m spectrum to measure
the object's spectral type. In Section~\ref{midir} we examine the
spectral energy distribution (SED) of \object\ over the range
$1-8$\,$\mu$m, using photometry in seven bands, to obtain an empirical
estimate of its temperature. In Section~\ref{fits} the near-infrared
spectrum is compared to theoretical spectral models, providing
estimates of the source's fundamental physical properties, effective
temperature, surface gravity, and metallicity, and thereby mass, and
age, through the use of evolutionary models. In Section~\ref{ammonia}
we report the results of higher-resolution spectroscopy in the {\em H}
band, to search for NH$_3$ absorption. Section~\ref{summ} provides a
brief summary of our results.

The UKIDSS {\zyjhk} photometry is based on Vega \citep{he06}, and all
magnitudes quoted are in the Vega system unless explicitly stated as
AB magnitudes (denoted by $i(AB)$, $z(AB)$, etc.).  Conversions
between AB and Vega for the SDSS {\em ugriz} and UKIDSS {\zyjhk} bands
are given by \cite{he06}.

%%%%%%%%%%%%%%%%%%%%%%%%%%%%%%%%%%%%%%%%%%%%%%%%%%%%%%%%%%%%%%%%%%%%%%%%%%%%%%

\section{Observations and data reduction}
\label{obs}

\object\ was identified as a candidate T dwarf after a largely
automated search of the UKIDSS DR1 database (Section~\ref{ident}),
after which follow-up photometry confirmed that its colours are
consistent with that interpretation (Section~\ref{photom}). Photometry
at mid-infrared wavelengths has been obtained
(Section~\ref{spitzerphot}) and the proper motion has been measured
using a one-year baseline (Section~\ref{astromsection}).
Near-infrared spectroscopy reveals \object\ to be a brown dwarf of
very-late spectral type (Section~\ref{spectro}).

\subsection{Identification}
\label{ident}

\object\ was discovered as a product of a general search of the UKIDSS
DR1 catalogue for objects with extreme optical-to-near-infrared
colours, including very cool brown dwarfs, as well as quasars of
redshift $z>6$ \citep{he06}. The selection was made by a SQL query of
the WFCAM Science Archive (WSA). A detailed description of the WSA,
together with examples of SQL queries, is provided by \citet{hambly}.
The first step in the search was to match
the UKIDSS DR1 catalogue to SDSS DR5 \citep{adel}, retaining both
unambiguous matches (within a match radius of 2\farcs0) with $i - Y
\geq 3$, as well as UKIDSS sources with no SDSS counterpart.  The
sample was further restricted to point-like sources with $Y \leq
19.5$.  The sample was cleaned of cross-talk\footnote{Saturated images
  of bright stars observed by WFCAM produce electronic ghost images in
  nearby detector channels, at integer multiples of the detector
  channel spacing of 128 pixels. The phenomenon, known as cross-talk,
  is illustrated and described in more detail in \citet{dye06}.
  Because ghosts appear at the same location in each of the UKIDSS
  filters they are often catalogued as being detected in multiple
  bands, but, since they are not real, they have no SDSS counterparts,
  thus masquerading as potentially interesting red sources.}
artefacts by rejecting sources located up to seven integer multiples
of 51\farcs2 \ in RA or dec from bright ($J \leq 13.5$) 2MASS stars.

The UKIDSS DR1 includes LAS fields covering some $100 \, {\rm{deg}}^2$
within a section of the multiply-scanned SDSS Stripe 82 \citep[the
  $\sim$212.5\,deg$^2$ defined by
  $-25^{\circ}<\alpha(2000)<+60^{\circ}$,
  $-1.25^{\circ}<\delta(2000)<+1.25^{\circ}$,][]{stough}. Therefore it
was possible to obtain improved $i$- and $z$-band photometry of
candidates within these fields \citep[as described
  in][]{venemans}. Most of the candidates were revealed to have
colours consistent with their being M dwarfs and were thus rejected.

\begin{figure}
\includegraphics[width=8cm]{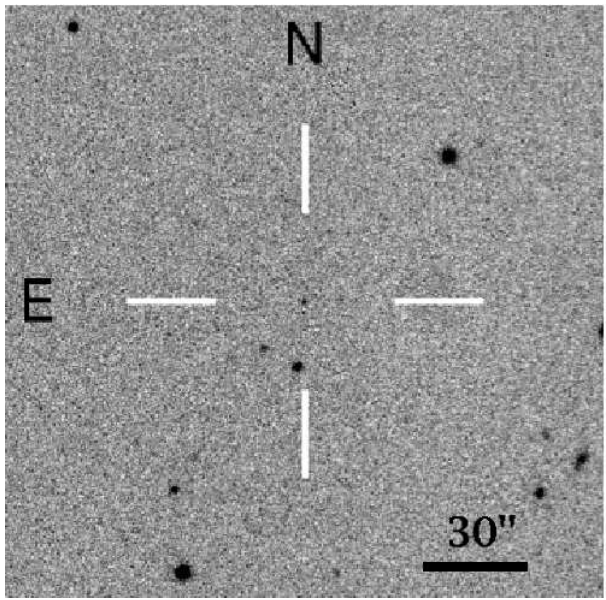}
\caption{The UKIDSS DR1 $Y$-band image of ULAS
  J003402.77$-$005206.7. The exposure time is 40s. The field of view
is $3\arcmin$ on a side.}
\label{fchart}
\end{figure}

The resultant list of red point--sources could include both
high-redshift quasars and brown dwarfs. We used the synthetic colours
of \citet{he06} to divide the list into a sample of candidate quasars,
with $\yj \leq 0.8$, and a sample of candidate brown dwarfs, with
$\yj > 0.8$. Quasars of redshifts $6<z<7$ have predicted colours
\yj$\sim0.5$, \jh$\sim0.4$, while mid to late T dwarfs have predicted
colours \yj$\sim1.1$, $-0.5<\jh<0.3$ \citep{he06}. \object\ was
located in the sample of quasar candidates, and stood out as the only
candidate which remained undetected in any band in the co-added SDSS
images.

The $Y$-band discovery image of \object\ is shown in
Fig.~\ref{fchart}.  The UKIDSS DR1 point--source photometry
\citep[{\tt apermag3},][]{dye06} is $Y=18.90\pm0.10$,
$J=18.14\pm0.08$, $H=18.50\pm0.22$, $K>17.94$ (5$\sigma$ detection
limit, established from the measured sky noise). The measured colours
\yj$=0.76\pm0.13$, \jh$=-0.36\pm0.23$, were ambiguous, the \yj\ colour
suggesting a quasar, and the \jh\ colour suggesting a brown dwarf, and
so a number of follow-up observations were made.

\subsection{Optical and near-infrared photometry}
\label{photom}

To clarify the nature of the source, higher S/N images in the $z$,
$J$, $H$, and $K$ bands were obtained.  The best photometry for the
source across the wavelength range $z$ to $K$ is summarised in
Table~\ref{tab_photom}. Some additional photometry, subsequently
superseded, is provided in the text, below.

The source was observed for 1800\,s with filter $\#611$ in EMMI, at
the NTT, on the night beginning 2006 August 18. This filter/detector
combination is referred to here as $z_n$ and is similar to the SDSS
$z$ band. To calibrate the image we first used the measured CCD
sensitivity curve and the filter transmission curve to establish the
colour relation between $z_n$ and SDSS $i$ and $z$. Briefly, the AB
magnitudes in the $i$, $z$, and $z_n$ bands of dwarf stars in the Gunn
Stryker atlas \citep{gs83} were computed by appropriate integration
under the bandpasses, to establish the colour relation
$z_n(AB)=z(AB)-0.05(i(AB)-z(AB))$ for dwarf stars from O to
M. Photometry in $i(AB)$ and $z(AB)$ of SDSS DR5 sources in the field
was then converted to $z_n(AB)$, and the brightness of the source was
measured by relative photometry using a fixed aperture, and is
recorded in Table \ref{tab_photom}. We also computed the colours of
some late T dwarfs with good spectrophotometry, to determine that
\object\ would be about 0.2 mag. fainter in $z(AB)$.

The offset to the Vega system is $z_n(AB) =z_n+0.563$, computed by
integrating the spectrum of Vega of \citet{bohlin} under the $V$ and
$z_n$ bandpasses, adopting $V=+0.026$, and demanding $z_n=V$ \ie, zero
colour for Vega.

\begin{table}
\centering
\caption{Optical and near-infrared photometry of the source \object.}
\begin{tabular}{@{}lccc@{}}
\hline
 Band      & mag.           & Instrument &  Date      \\ 
           &                &            &    \\ \hline
 $z_n(AB)$ & $21.91\pm0.05$ & EMMI       & 2006--08--18 \\
 $Y$       & $18.90\pm0.10$ & WFCAM      & 2005--10--04 \\
 $J$       & $18.15\pm0.03$ & LIRIS      & 2007--01--03 \\
 $H$       & $18.49\pm0.04$ & LIRIS      & 2007--01--03 \\
 $K$       & $18.48\pm0.05$ & UFTI       & 2006--09--02 \\ \hline
\end{tabular}

\label{tab_photom}
\end{table}

The source was observed with the UFTI instrument \citep{roche} on
UKIRT in $J$ and $H$ for 450\,s each, on the night beginning 2006
August 7, and in $K$ for 1620\,s, on the night beginning 2006
September 2. The data were calibrated using the UKIRT Faint Standards
\citep{leg06}. The filters are from the Mauna Kea Observatories set
\citep[MKO;][]{tokunaga}, and the photometric system is virtually
identical to the WFCAM system \citep{leg06}. The data were reduced
using the astronomical imaging pipeline ORAC-DR
\citep{cav03}.\footnote{The ORAC-DR Imaging Data Reduction User Guide
  is available at
  http://www.starlink.ac.uk/star/docs/sun232.htx/sun232.html.} The
result for $K$ is provided in Table \ref{tab_photom}. The other
measures were $J=18.14\pm0.03$, and $H=18.51\pm0.05$. This are not
provided in the table as they have been superseded (see below).

The blue colour $J-H=-0.37\pm0.06$, measured with UFTI, established
that the source is a cool brown dwarf rather than a quasar. It was
possible to estimate an approximate spectral type by comparing the
measured colours $z_n(AB)-J=3.77\pm0.06$, $J-H=-0.37\pm0.06$, with the
table of synthetic colours and spectral types of T dwarfs from
\citet{he06}. We first convert $z_n(AB)$ to $z(AB)$ for \object\ by
adding 0.2 (as above), and then convert to the Vega system by
subtracting the AB offset of 0.533, from Table 4 in \citet{he06},
yielding the approximate Vega colour $z-J=3.44\pm0.06$. Taking the
colours and spectral types of T dwarfs from Table 10 of \citet{he06},
we found there is a correlation between each of the two colours and
spectral type. From the measured colours of \object, the errors, and
the scatter in the relations, we estimated an approximate spectral
type T$7\pm1$. Accordingly the object was ranked as a high priority
candidate for spectroscopy.

The target was reobserved at higher S/N in $J$ and $H$ with the LIRIS instrument
\citep{acosta} on the WHT, on the night of 2007 January 3. Integration
times were 900\,s and 1000\,s respectively. These data were calibrated
to the UKIDSS DR1 images using nearby bright unsaturated stars. We
added an error of 0.02\,mag. in quadrature to the random photometric
errors, to account for the accuracy of the UKIDSS calibration. The
results are provided in Table \ref{tab_photom}.

The colour $Y-J=0.75\pm0.10$ of \object\ is rather blue in comparison
with other T dwarfs found in UKIDSS. The average colour of the nine
new spectroscopically confirmed T4 to T7.5 dwarfs listed in
\citet{kendall} and \citet{lodieu} is 1.13. For completeness we note
that the zero point of the UKIDSS {\em Y} photometry is under review.
Preliminary analysis indicates that a constant 0.09\,mag. should be
added to the UKIDSS DR1 {\em Y} values, to place the photometry on the
Vega system \citep{wa07a}. Nevertheless, recalibration will not, of
course, alter the difference in colours quoted above.

\subsection{Mid-infrared photometry}
\label{spitzerphot}

We obtained IRAC four-channel (3.55, 4.49, 5.73 and 7.87\,$\mu$m)
photometry of the source on UT date 2006 December 26.  The data were
obtained as part of the {\em Spitzer Space Telescope} DDT program
$\#281$. All four channels have 256$\times$256-pixel detectors with a
pixel size of 1\farcs2$\times$1\farcs2, yielding a
5\farcm2$\times$5\farcm2 field of view.  Two adjacent fields are
imaged in pairs (channels 1 and 3; channels 2 and 4) using dichroic
beam splitters.  The telescope is then nodded to image a target in all
four channels.\footnote {For more information about IRAC, see
  \citet{faz04} and the IRAC Users Manual at
  http://ssc.spitzer.caltech.edu/irac/descrip.html} We used exposure
times of 30~s and a 5-position medium-sized (52 pixels) dither pattern
repeated ten times, for a total observing time of 57.6 minutes.

The data were reduced using the post-basic-calibration data mosaics
generated by version 15 of the IRAC pipeline.\footnote{Information
  about the IRAC pipeline and data products can be found at
  http://ssc.spitzer.caltech.edu/irac/dh/} The mosaics were
flat-fielded and flux-calibrated using super-flats and global primary
and secondary standards observed by {\em Spitzer}.

We performed aperture photometry using an aperture with a 2-pixel (or
2\farcs4) radius, to minimise the contribution from a faint source
$3\farcs5$ to the south-east. To assess the influence of the
neighbouring source, we repeated the photometry using an aperture with
a 4-pixel radius. After applying the appropriate aperture
corrections, the results for the larger aperture are total fluxes 6 to
0\,per cent brighter, for channels 1 through 4,
respectively. Therefore contamination of the 2-pixel photometry by the
neighbouring source will be at a level substantially below these
values. To convert the 2-pixel fluxes to total fluxes, we applied
aperture corrections as described in Chapter~5 of the IRAC Data
Handbook$^3$ of 1.205, 1.221, 1.363 and 1.571 to channels 1 through 4,
respectively. The photometry was converted from milli-Janskys to
magnitudes on the Vega system using the zero-magnitude fluxes given in
the IRAC Data Handbook (280.9, 179.7, 115.0, and 64.1 Jy for channels
1 to 4, respectively).  Note that IRAC observations are reported as a
flux density at the nominal wavelengths given in Table \ref{spitzer},
assuming that the target has a flux density $f_{\nu}\propto 1/\nu$.
This assumption is not valid for T dwarfs and so the results given in
the Table should not be used to derive a spectral flux at the nominal
wavelength.  However, if the mid-infrared spectral energy distribution
is known, \citet{cush06} show how the IRAC fluxes can be used to
photometrically calibrate the spectrum \---\ or compute the `color
correction', in the terminology of the IRAC Data Handbook.

Photometric random errors were derived from the uncertainty images
that are provided with the post-basic-calibration data. The magnitudes
and errors are given in Table \ref{spitzer}. There are, in addition,
systematic errors from a variety of sources \citep{reach}, and the
IRAC Data Handbook recommends citing $5\%$ uncertainty for the
absolute calibration. This includes a contribution from the
calculation of the `color correction', which therefore does not apply
here. Additional sources of uncertainty include: the accuracy of the
absolute calibration \citep[estimated to be $2\%$ by][]{reach},
systematic errors introduced by pipeline dependencies
\citep[of comparable size,][]{leg07a}, the variable pixel scale
across the field of view, and the variation of the quantum efficiency
across a pixel. For observations confined to near the centre of the
array, the last two effects combined appear to be at a level of $<2\%$
in an individual exposure, as measured from the scatter in repeat
dithered observations \citep[Fig. 3 in][]{patten}, and therefore
extremely small after averaging dithered observations. Accordingly, we
adopt the total photometric uncertainty to be the sum in quadrature of
the values given in Table \ref{spitzer} plus 3\,per cent.

\begin{table}
\centering
\caption{Mid-infrared (Vega) photometry of the source \object. A 3\,per cent
error should be added in quadrature to the quoted random errors to
account for systematics.}
\begin{tabular}{@{}cc@{}}
\hline
 Band      & mag.           \\ 
 $\mu$m    &                \\ \hline
 3.55      & $16.28\pm0.01$ \\
 4.49      & $14.49\pm0.01$ \\
 5.73      & $14.82\pm0.04$ \\
 7.87      & $13.91\pm0.05$ \\ \hline
\end{tabular}

\label{spitzer}
\end{table}

\subsection{Astrometry}
\label{astromsection} 

The original DR1 {\em YJHK} images were all taken on the night
beginning 2005 October 4, and serve as a first epoch for the
measurement of \object's proper motion. Since the $J$ image has the
highest S/N this is the best wavelength for a second--epoch image.
The source name ULAS J003402.77$-$005206.7 is derived, following the
protocol of the International Astronomical Union, from the UKIDSS DR1
coordinates measured on the {\em Y} image of $0^{\mbox {\scriptsize
    h}}34^{\mbox {\scriptsize m}}2.771^{\mbox {\scriptsize s}}
-0^{\circ}52\arcmin6.78\arcsec$. The coordinates are calibrated to
2MASS, and are on the International Celestial Reference System. The
absolute accuracy is better than 100\,mas on each axis \citep{dye06}.

A deeper WFCAM {\em J} image was obtained on the night beginning 2006
December 4. The source was observed for 400\,s, and was located on the
same detector as for the first epoch, in order to minimise systematic
errors due to distortion. The measured shifts in RA and dec relative
to the first epoch, and the computed proper motion are provided in
Table \ref{astromtable}. We detect a significant proper motion of
$0\farcs37\pm0\farcs07$/yr.

The LIRIS $H$-band image (Section \ref{photom}), compared to the first
epoch {\em H} image, provides an independent measure of the proper
motion. The result, provided in the second line of Table
\ref{astromtable}, is consistent with the $J$-band measure, but with a
larger uncertainty. Rather than combine the two results, we retain the
$J$-band measure as the best estimate, because the differential
systematic errors due to distortion should be smaller, since the two
$J$-band images were taken with the same instrument.

\begin{table*}
\centering
\caption{Measured shifts relative to the base epoch of 2005--10--04, and
  corresponding proper motions. The proper motions (but not the
  shifts) have been corrected for parallax, assuming a distance of 18\,pc.}
\begin{tabular}{@{}ccccccc@{}}
\hline
 Band      & epoch 1 & epoch 2 & $\Delta \alpha$ & $\Delta \delta$ & 
  $\mu_\alpha$ & $\mu_\delta$  \\  
     &  &  & $\arcsec$ & $\arcsec$ & $\arcsec$/yr &
 $\arcsec$/yr \\ \hline 
 $J$ & 2005--10--04 & 2006--12--04 & $-0.18$ & $-0.43$ & $-0.12\pm0.05$ &
 $-0.35\pm0.05$ \\
 $H$ & 2005--10--04 & 2007--01--03 & $-0.09$ & $-0.39$ & $-0.04\pm0.07$ &
 $-0.31\pm0.07$ \\ \hline
\end{tabular}

\label{astromtable}
\end{table*}

\subsection{Spectroscopy}
\label{spectro}

\begin{figure*}
\includegraphics[width=17cm]{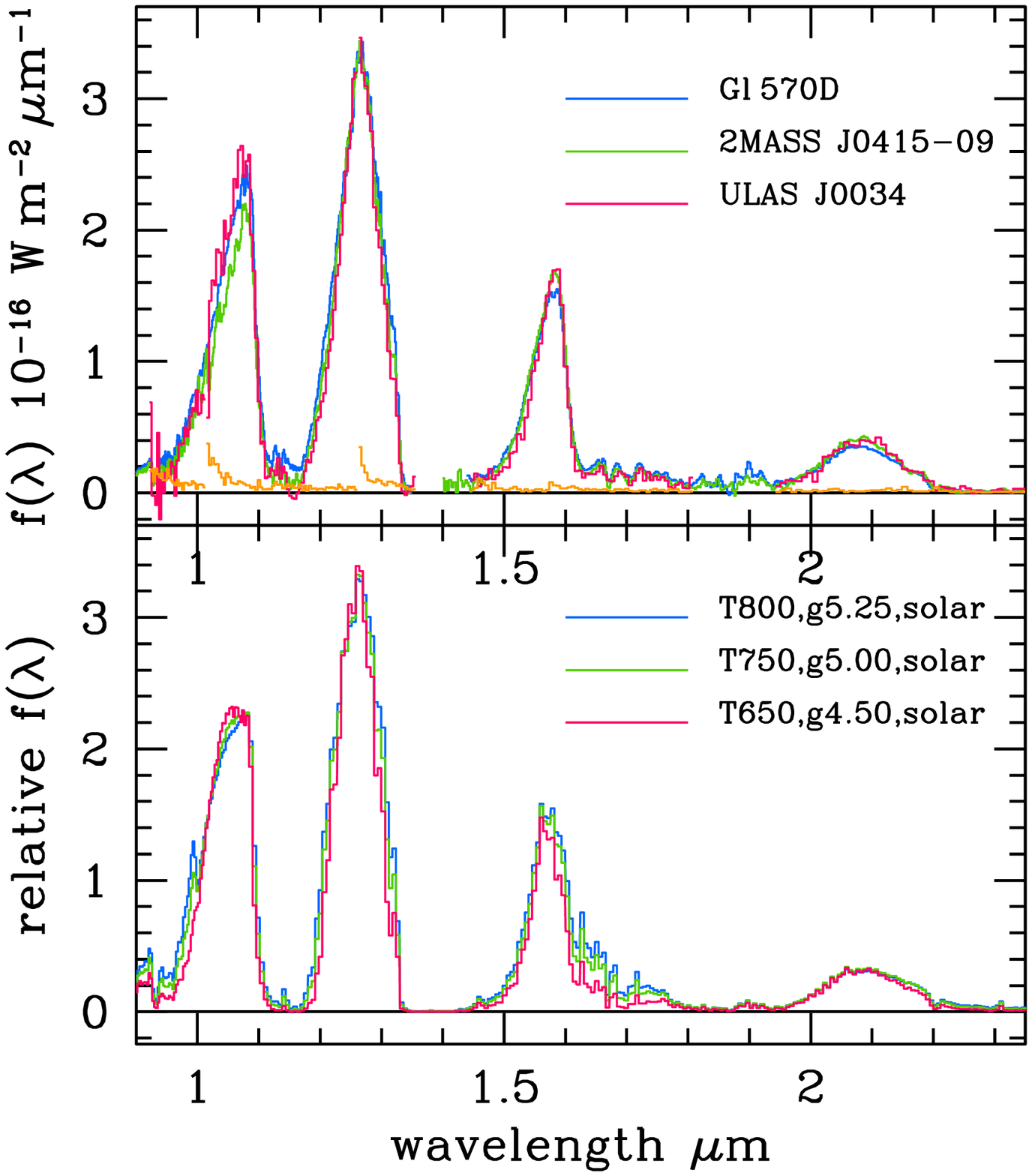}
\caption{{\em Upper:} GNIRS spectrum of \object, binned by a factor
  12, plotted magenta. The orange line plots the 1$\sigma$ error
  spectrum. The five separate sections correspond to different
  spectral orders.  The blue and green lines plot the spectra of the
  T7.5 brown dwarf \gobject\ and the T8 brown dwarf \mobject, for
  comparison. These spectra were scaled to the crown of the $J$-band
  peak of \object, over the wavelength range 1.25--1.28\,$\mu$m.  {\em
    Lower:} Solar--metallicity model spectra representing
  \object\ (magenta), \gobject\ (blue), and \mobject\ (green).}
\label{gnirsfull}
\end{figure*}

\begin{figure*}
\includegraphics[width=17cm]{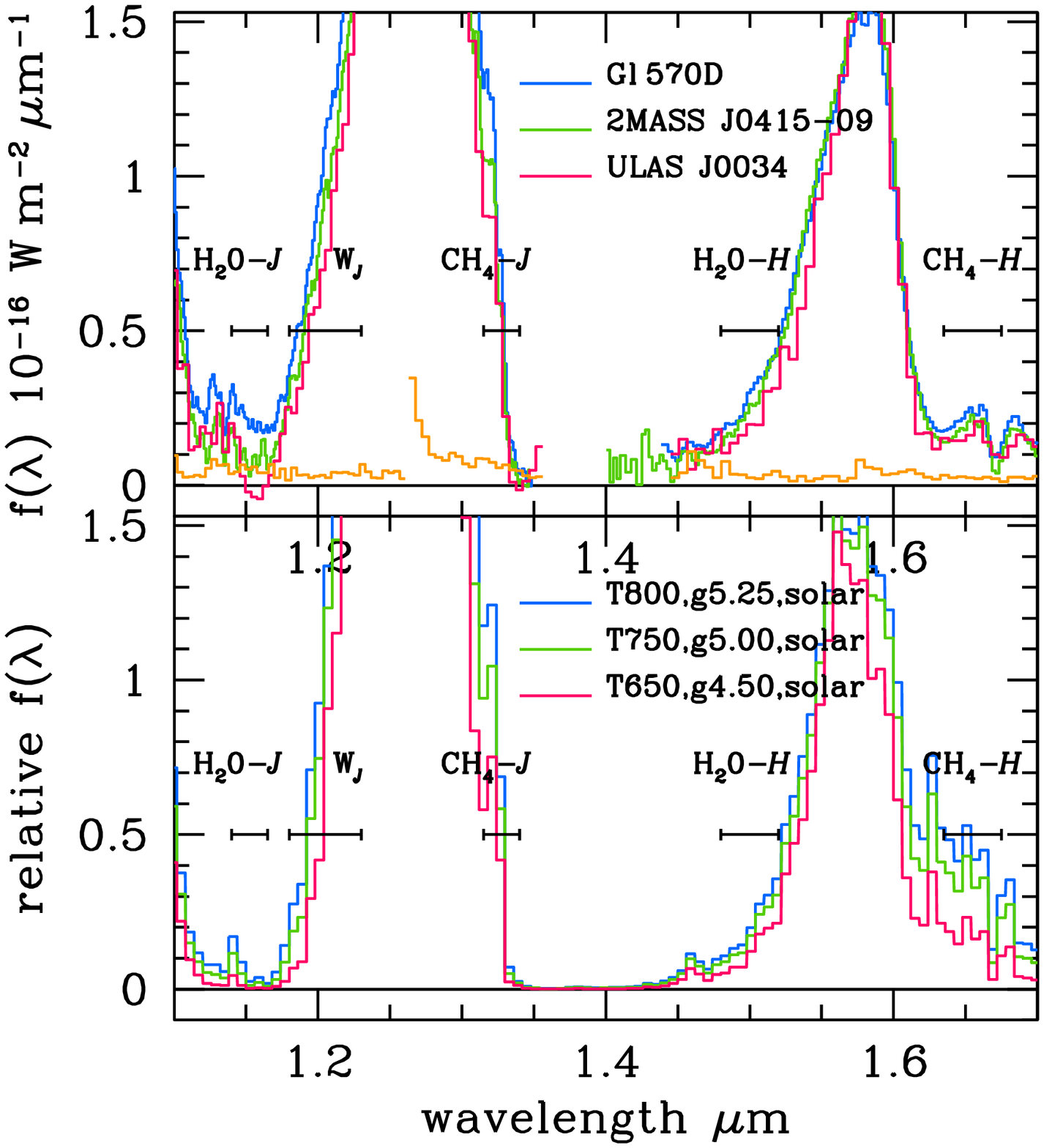}
\caption{Zoomed portion of Fig.~\ref{gnirsfull} showing the absorption
  troughs either size of the {\em J} and {\em H} peaks. Wavelength
  regions of the spectral classification bands (Table
  \ref{tab_indices}) are marked.  {\em Upper:} GNIRS spectrum of
  \object, binned by a factor of 12, plotted magenta. The orange line
  plots the 1$\sigma$ error spectrum. The separate sections correspond
  to different spectral orders.  The blue and green lines plot the
  spectra of the T7.5 brown dwarf \gobject\ and the T8 brown dwarf
  \mobject, for comparison. These spectra were scaled to the crown of
  the $J$-band peak of \object, over the wavelength range
  1.25--1.28\,$\mu$m. {\em Lower:} Solar--metallicity model spectra
  representing \object\ (magenta), \gobject\ (blue), and
  \mobject\ (green).}
\label{gnirszoom}
\end{figure*}

\subsubsection{GNIRS low-resolution spectrum}

The Gemini Near-Infrared Spectrograph \citep[GNIRS;][]{gnirs} on
Gemini South was used to make quick response observations of \object,
through program GS-2006B-Q-36. GNIRS was used in cross-dispersed mode
with the 32\,l/mm grism, the $1\farcs0$ slit and the short camera, to
obtain 0.9--2.5\,$\mu$m R$\sim$500 spectra. Two triggers were made on
the target, providing a shorter and longer total integration time of
16 and 58.7 minutes, on the nights of 2006 August 26 and 2006 Oct 1,
respectively. The target was nodded 3\farcs 0 along the slit in an
``ABBA'' pattern using individual exposure times of 240\,s and 220\,s
respectively. Calibrations were achieved using lamps in the
on-telescope calibration unit. Early F stars were observed as telluric
standards, at an airmass that matched the mid-point airmass of the
target observation.

Data reduction was initially undertaken with tasks in the Gemini GNIRS
IRAF Package. The cross-dispersed spectrum comprises five highly
curved spectra, spatially separated, that correspond to different
grating orders over the wavelength of interest. The reduction
procedure included correction for offset bias, order separation,
median stacking at the A and B positions, sky subtraction,
S-distortion correction (\ie\ straightening the curved spectra), and
wavelength calibration. The data were not flat-fielded. Apart from bad
pixels, the variation of the array quantum efficiency is sufficiently
small that flat-field correction makes no discernible reduction in the
noise in the spectra. Further reduction was carried out in IDL. Sky
residuals were fit, and subtracted, using a surface constructed via a
series of linear fits across the slit (excluding pixels within the
spectral apertures). Spectra were then extracted within $1\farcs5$
apertures and summed for the A and B positions. Noise spectra were
also extracted. The spectra were sigma-clipped and calibrated onto a
relative flux scale using the spectrum of the telluric standard (after
appropriate interpolation across any hydrogen absorption lines) and a
black--body function for \tf$=7500$\,K. The spectral orders were then
trimmed of their noisiest portions, and the spectrum was scaled to
absolute flux by the $J$-band photometry. The final spectrum consists
of an average of the longer and shorter triggered observations,
weighted by exposure time.

The pixel scale oversamples relative to critical sampling by a factor
three (the slit is 6.7 pixels wide). Because the spectrum S/N is
relatively low, and bearing in mind that the principal diagnostic
features are broad, for the purposes of display the spectrum has been
binned by a factor of 12 (\ie\ two resolution elements per pixel).
The complete spectrum is shown as the magenta line in the upper panel
of Fig.~\ref{gnirsfull}, where it is compared to the spectra of two
well-known very cool brown dwarfs, the T7.5 dwarf \gobject, discovered
by \citet{bur00}, and the T8 dwarf \mobject, discovered by
\citet{bur02}. The comparison spectra are from \citet{geb01} and
\citet{kna04}, respectively. The comparison spectra are of comparable
resolution to the GNIRS spectrum. The lower panel plots model spectra,
and is discussed in Section \ref{speccomp}.  Fig.~\ref{gnirszoom}
provides a zoom of the absorption troughs either side of the {\em J}
and {\em H} peaks.

We synthesised colours from the spectrum, following the procedure
described by \citet{he06}. Briefly, this involves integration of the
spectrum under the passband response curves, with appropriate
normalisation by the same integration performed on the spectrum of
Vega. The resulting colours are as follows, and are in satisfactory
agreement with the measured values, provided in parentheses:
$Y-J=0.79\pm0.03\,(0.75\pm0.10)$, $J-H=-0.46\pm0.01\,(-0.34\pm0.05)$,
$H-K=0.05\pm0.02\,(0.01\pm0.07)$. As previously noted the measured
$Y-J$ colour is unusually blue. The synthetic $Y-J$ colour
corroborates this finding. The synthetic $Y-J$ colours of
\gobject\ and \mobject\ are $0.95$ and $1.06$ respectively
\citep{he06}.

\subsubsection{ISAAC intermediate-resolution spectrum}

An intermediate-resolution spectrum covering the wavelength range
$1.50-1.58\mu$m was obtained with the ISAAC instrument on the ESO VLT
using Director's Discretionary Time, under programme
278.C-5014(A). The purpose of the spectrum was to search for
additional absorption lines in comparison with the spectrum of
\mobject, that could explain the excess absorption seen in the
low-resolution spectrum of \object\ (Figs \ref{gnirsfull} and
\ref{gnirszoom}), and might be due to NH$_3$. We used the MR grating
and a $1\arcsec$ slit, providing a nominal resolving power of
$\sim3000$. The average seeing was $0\farcs7$, meaning that the actual
resolving power was nearer 4000. The dispersion was 0.807\AA,
corresponding to $\sim5$ pix per resolution element.

The source was observed on the nights beginning 2007 Jan 3 and 11. On
each night, integrations of 600s each were obtained at four separate
slit positions, providing a total of 80min on source. As described in
\citet{weath} higher S/N can be achieved with multiple slit positions,
compared to the traditional method of observing in ABBA sequence and
subtracting in pairs. For each slit position a sky frame was created
by averaging the three other frames from the same night, and was
subtracted. The data were then flat-fielded, and improved sky
subtraction achieved by fitting a function up each line. All eight
frames were then registered to the nearest pixel spatially and
spectrally, scaled to a common count level, and averaged, ignoring bad
pixels. Wavelength calibration was achieved using the sky lines, and
was found to be linear to better than a pixel. Therefore rebinning was
not required in applying the wavelength solution. Observations of
standard stars were used to correct for atmospheric absorption
(everywhere less than 10 per cent over this wavelength region), and to
calibrate onto a relative flux scale. Absolute calibration was
achieved by scaling to the GNIRS spectrum.

The spectrum is discussed later in Section \ref{ammonia}, where the
flux-calibrated spectrum, together with the error spectrum, is plotted
in Fig.~\ref{isaac}.

\begin{table*}
\centering
\caption{Measured values of the spectral classification indices of
  \citet{bur06a}, for the source \object.
  Each index is labelled by both the absorption species and the
  band in which it appears, and is defined by the ratio of the summed
  flux in the two wavelength ranges given.
  The error associated with the measured index is only the 
  random contribution obtained by propagating the uncertainties implied
  by the noise spectrum.  For each index the range associated with the 
  T8 classification is given, along with the classification for 
  \object\ implied by the measured value.}
\begin{tabular}{@{}lllccl@{}}
\hline
 Band     & numerator   & denominator & value & T8    & type \\
          & wavelength  & wavelength  &       & range &      \\
      &\multicolumn{1}{c}{$\mu$m}&\multicolumn{1}{c}{$\mu$m}&  &  \\ \hline
H$_2$O$-J$ & 1.14 $-$1.165 & 1.26$-$1.285 & $0.012\pm0.006$ & $0.02-0.07$ & T8.5 \\
CH$_4$$-J$ & 1.315$-$1.34  & 1.26$-$1.285 & $0.144\pm0.009$ & $0.15-0.21$ & T8.5 \\
H$_2$O$-H$ & 1.48 $-$1.52  & 1.56$-$1.6   & $0.133\pm0.010$ & $0.14-0.20$ & T8.5 \\
CH$_4$$-H$ & 1.635$-$1.675 & 1.56$-$1.6   & $0.096\pm0.006$ & $0.07-0.15$ & T8   \\
CH$_4$$-K$ & 2.215$-$2.255 & 2.08$-$2.12  & $0.091\pm0.015$ & $. . .$ & $\geq$T7 \\ \hline

\end{tabular}

\label{tab_indices}
\end{table*}

\section{Spectral type}
\label{spectype}

To determine the spectral type we follow the revised classification
scheme for T dwarfs of \citet{bur06a} which unifies and refines the
two independent preliminary classification schemes of \citet{geb02}
and \citet{bur02}, both of which are grounded in the methodology of
\citet{mk}.  Classification is based on the depth of the absorption
bands due to H$_2$O and CH$_4$ in the near-infrared, between the {\em
  Y}, {\em J}, {\em H}, and {\em K} peaks, which strengthen towards
later spectral types, and are nearly saturated by T8. Classification
is achieved by comparison of the object spectrum against a set of
template spectra, or from the measured values of a set of spectral
indices that quantify the strength of the absorption.

In Figs. \ref{gnirsfull} and \ref{gnirszoom} the spectrum of
\object\ is compared against that of the T8 spectral template,
\mobject.  Also plotted is the spectrum of the well studied brown
dwarf \gobject, classified as T7.5 by \citet{bur06a}. The three
spectra have been scaled to the crown of the {\em J-}band peak. There
is visibly a progressive narrowing of the {\em J-}band peak in moving
from \gobject, to \mobject, to \object, as well as successively
stronger absorption in the H$_2$O trough near $1.15$\,$\mu$m.  If the
spectra were instead normalised to the {\em H-}band peak the same
sequence would be evident in the blue wing of the peak, near
$1.5$\,$\mu$m. This visual comparison rules out a spectral type as
early as T7.5 for \object, and suggests a classification of T8 or
T8.5.

The $Y$-band peak is relatively stronger in both \object\ and
\gobject, compared to \mobject, but this region is not used in current
spectral classification schemes. A weak absorption feature in the
spectrum of \object\ near 2.0\,$\mu$m coincides with a strong telluric
band, and is probably not intrinsic to the source.

Besides comparison against template spectra, we have also measured the
(primary) spectral indices defined by \citet{bur06a}, and these are
given in Table \ref{tab_indices}.  The first three indices imply a
classification of T8.5, the fourth gives T8 and the fifth, CH$_4$$-K$,
saturates at T7, and therefore is not useful for this source.  Taken
together the indices suggest that T8.5 is the appropriate class. This
is in agreement with the value from the template comparison, and
therefore is our adopted value.

Although the precision of the classification scheme itself is only 0.5
of a class, this is the first brown dwarf discovered with a
classification later than T8. For this reason we have investigated the
uncertainty in the class in a formal way, treating class as a
continuous variable. Then the value of each index and its uncertainty,
together with the mean and range of the parameter defining the T8
class, provide four estimates of the parameter. An unweighted mean of
the four estimates yields a class T8.5$\pm$0.1 (error on the
mean). For the weighted mean the uncertainties on the indices were
doubled, to force $\chi^2=3$ (for three degrees of freedom), which led
to the result T8.4$\pm$0.1. A systematic error in the sky subtraction,
summed over any band, could make a larger contribution to the error
budget. Nevertheless this analysis supports the result from the
template comparison that the class cannot be earlier than T8.

\begin{figure}
\includegraphics[width=8.2cm]{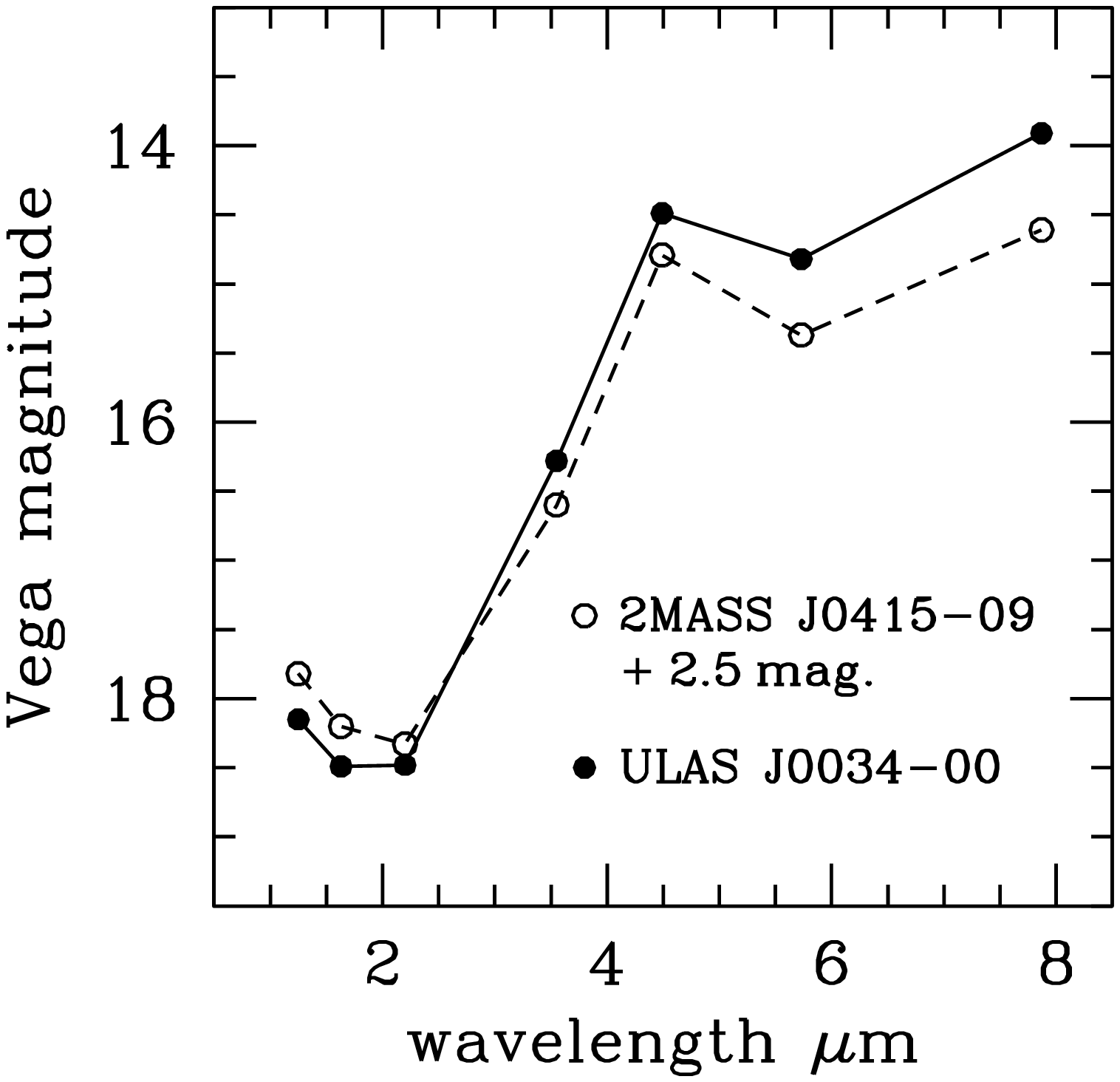}
\includegraphics[width=8.2cm]{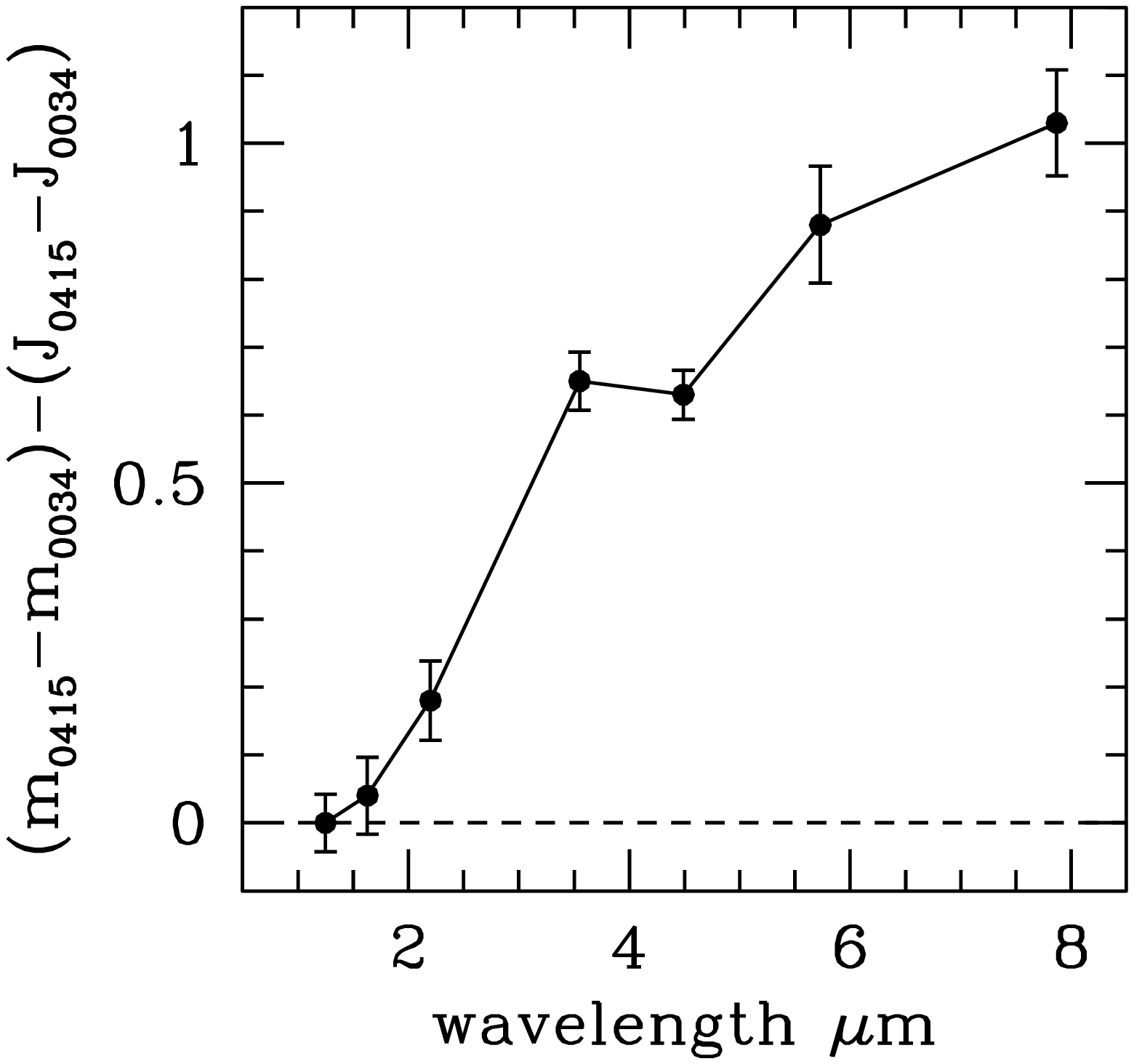}
\caption{{\em Upper:} Comparison of 1--8\,$\mu$m SEDs of \object\ and \mobject.
  The data for the latter object have been shifted by
  $+2.5\,$mag. so that the SEDs overlap. Uncertainties are not plotted
  as they are comparable to the size of the symbols. {\em Lower:} The
  same information, plotted as the magnitude difference
  $m_{0415}-m_{0034}$ minus the magnitude difference in the {\em J}
  band, with uncertainties, showing that \object\ is 1\,mag. redder
  over this wavelength range.}
\label{sed}
\end{figure}

\begin{figure}
\includegraphics[width=8.2cm]{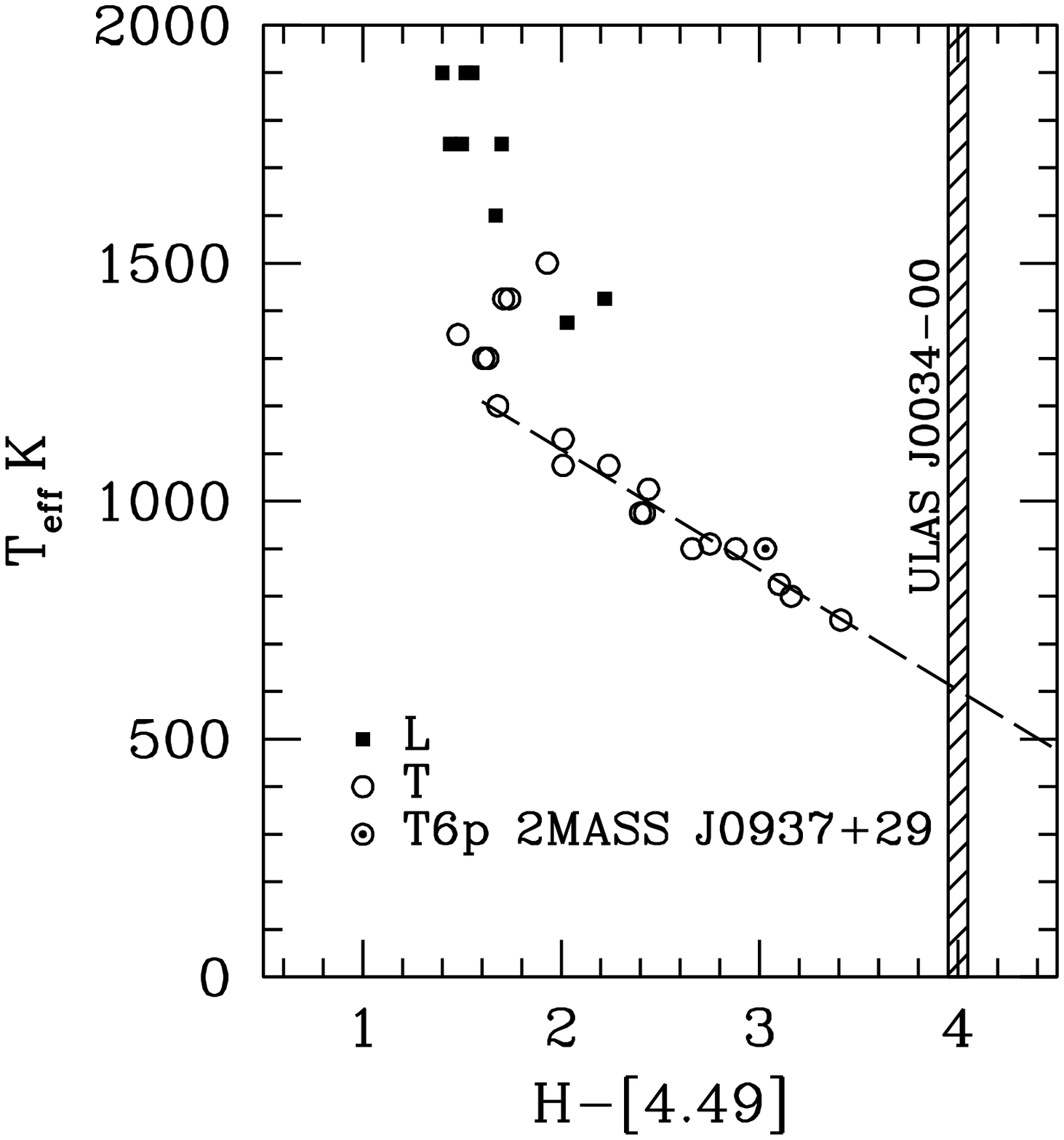}
\includegraphics[width=8.2cm]{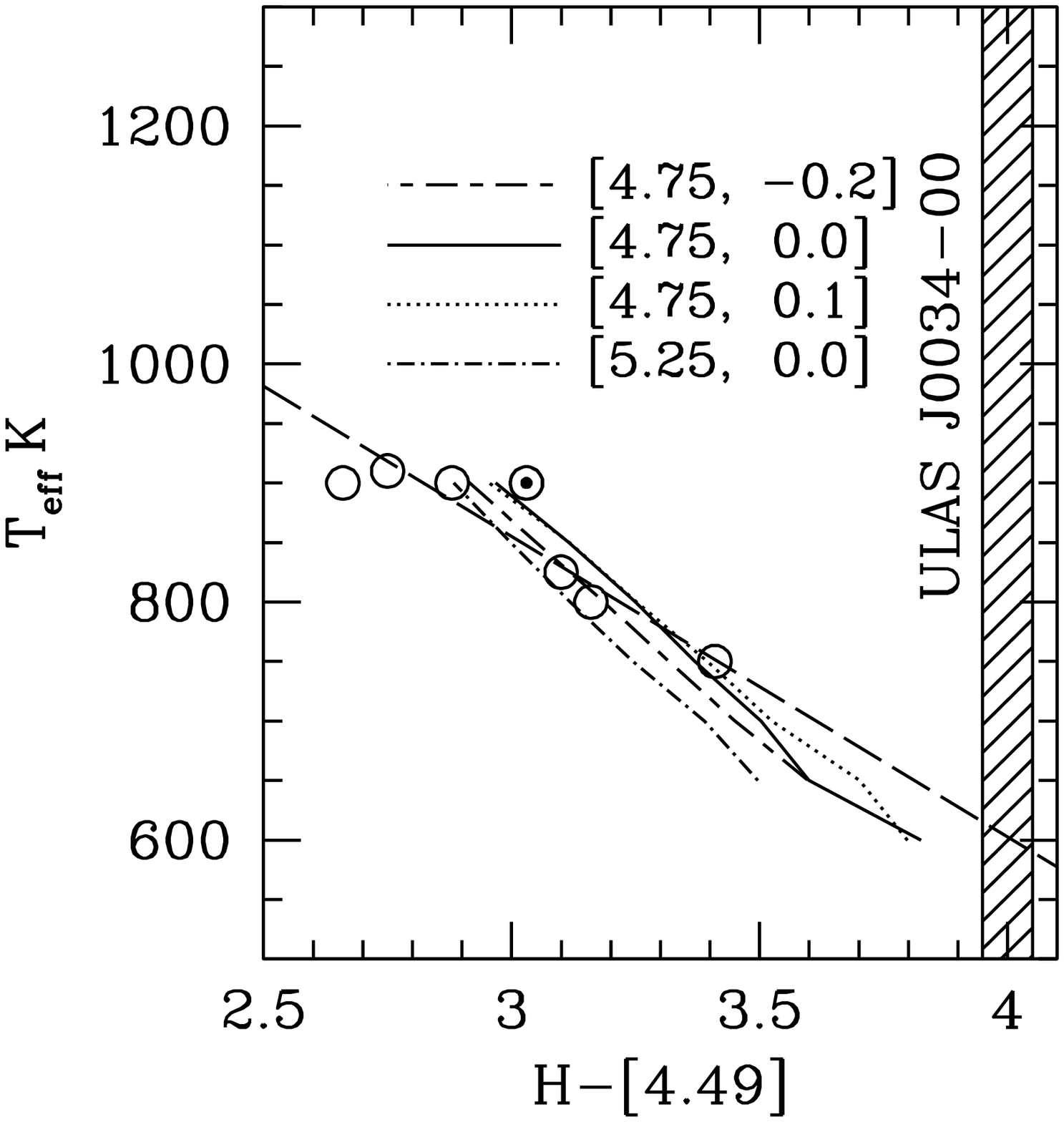}
\caption{{\em Upper:} Relation between \tf\ and $H-[4.49]$ colour for
  L dwarfs (filled squares) and T dwarfs (open circles) with
  well-determined temperatures. The dashed line is a linear fit to the T
  dwarfs with \tf$<1250$\,K. The T6p dwarf 2MASS J0937$+$29
  is discussed in the text. The colour of \object,
  $H-[4.49]=4.00\pm0.05$, is indicated by the shaded region. {\em
    Lower: } Zoom of the upper plot, with model \tf$-$colour tracks plotted
  for a range of gravity/metallicity combinations, listed in the
  key. The models are described in Section \ref{fits}. The long-dashed
  line is the fit from the upper plot.} 

\label{tempcol}
\end{figure}

\section{Effective temperature estimate from the spectral energy distribution
  from 1--8\,$\mu$m}
\label{midir}

The SED over the range 1--8\,$\mu$m, measured by the broad-band
photometry, provides further evidence of the extreme nature of
\object. \citet{patten} and \citet{leg07a} note relations between
various colours within this wavelength range and spectral type,
luminosity, and effective temperature.  \object\ is more extreme than
\mobject\ in nearly all the colours plotted.  In Fig.~\ref{sed} we
plot the photometry in the \jhk\, bands plus the four IRAC bands for
\object\ and for \mobject. For the latter the data were taken from
\citet{patten}, and we note that, as for \object, the \jhk\ photometry
is on the MKO system (no $Y$-band photometry for this source exists as
far as we are aware). The datapoints in Fig.~\ref{sed} should only be
interpreted in a relative sense because, as described in
Section~\ref{spitzerphot}, the IRAC magnitudes are not equivalent to
the corresponding flux at the nominal filter wavelength. The SED of
\object\ is significantly redder over the entire wavelength range
plotted, indicative of lower temperature. For example, colours for
\object\ (\mobject) are as follows;
$J-K=-0.33\pm0.06\,(-0.51\pm0.04)$,
$K-3.55=2.20\pm0.06\,(1.73\pm0.04)$,
$3.55-7.87=2.37\pm0.07\,(1.99\pm0.06)$.

The absorption bands of H$_2$O and CH$_4$ which define the T spectral
sequence are nearly saturated at T8. This means that spectral changes
in the near-infrared may be relatively insensitive to temperature for
objects cooler than \mobject.  Therefore it is conceivable that
\object\ is substantially cooler than \mobject, despite the rather
small near-infrared spectral differences. In order to estimate \tf\ we
have compiled a sample of L and T dwarfs that possess good temperature
estimates, as well as photometry in the seven bands from {\em J} to
[7.87] plotted in Fig.~\ref{sed}. Temperatures are taken from
\citet{golimowski04, luhman, leg07b}, and photometry from
\citet{patten,leg07a, luhman}.  We then plotted \tf\ against all
possible colour combinations, looking for the best correlation. We
avoided the {\em K}-band because the flux in this band is sensitive to
both gravity and metallicity at fixed temperature
\citep[e.g.][]{liulc}.

In Fig.~\ref{tempcol} we plot \tf\ against the {\em H}$-$[4.49] colour
for L and T dwarfs with \tf$<2000$\,K.  The L dwarfs become redder in
{\em H}$-$[4.49] towards cooler \tf, but there is a discontinuity in
the relation at the L/T transition, reflecting the strong spectral
changes that occur over a small temperature range. Nevertheless for
later T dwarfs with \tf$<1250$\,K there is a remarkably tight
correlation. The RMS deviation from the best-fit linear relation,
plotted in the figure, is only 28\,K. The sample plotted includes the
peculiar T6 dwarf 2MASS J0937+29. The {\em K-}band flux of this object
is strongly suppressed \citep{bur02}, due either to high gravity or
low metallicity or both \citep{bur06b}.  The object is the most
distant outlier in Fig.~\ref{tempcol}, located at \tf\ $=900$\,K, {\em
  H}$-$[4.49] $=3.00$. Nevertheless it lies only 52\,K above the line,
\ie, at $2\sigma$. This hints that for mid and late T dwarfs the {\em
  H}$-$[4.49] colour is relatively insensitive to variations in
gravity and metallicity.

Other than \object, the reddest object plotted in Fig.~\ref{tempcol}
is \mobject, which has $\mbox{{\em H}$-$[4.49]} = 3.41\pm0.04$ and an
effective temperature of $\tf = 750$\,K.  \object\ is considerably
redder, with {\em H}$-$[4.49] $=4.00\pm0.05$, implying that it is
correspondingly cooler.  Extrapolating the linear fit provides a
temperature estimate $\tf = 600$ $\pm 30$\,K for \object. The lower
panel in Fig.~\ref{tempcol} is a zoom of the region of interest. Also
shown are model tracks in this parameter space, computed from the
synthetic spectra described in Section~\ref{fits}, for a range of
gravities (parameterised by log\,$g$), and metallicities [m/H]. Over
the temperature range $700<\tf<900$\,K the model tracks fit the data
well, while the small scatter between the tracks implies that the
small scatter seen in the data is not simply fortuitous, but because
the $\mbox{{\em H}$-$[4.49]}$ colour is indeed relatively insensitive
to variations in metallicity and gravity. The model tracks remain
linear to cooler temperatures $\tf<700$\,K, providing some
justification for extrapolating the linear fit to the data. In fact
the tracks are somewhat steeper than the linear fit to the data, which
implies a temperature even lower than 600\,K. We defer a more detailed
discussion of the models in the mid-infrared region until completion
of scheduled mid-infrared spectroscopy.

Assuming the linear extrapolation of the colour--temperature
relationship to be valid, \object\ could have a comparable temperature
to \mobject\ only if it were a $\sim 5 \sigma$ outlier.  In this
respect it would be more extreme than 2MASS~J0937+29 (mentioned
above), and, if it was anomalous for the same reasons, its {\em K}-band
flux ought to be even more strongly suppressed, which is not seen.

\begin{figure*}
\includegraphics[width=17cm]{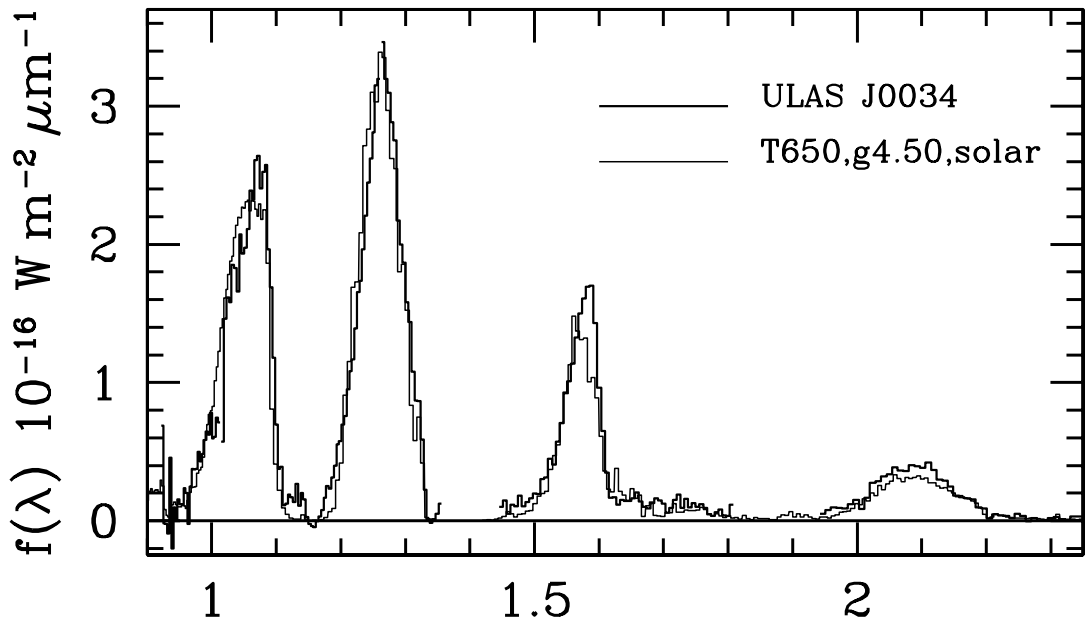}
\caption{Direct comparison of the best-fit model with the GNIRS
  spectrum, where the model has been scaled to the crown of the $J$-band peak.}
\label{datamodel}
\end{figure*}

\section{Fundamental parameters from spectral modelling}
\label{fits}

In this section we compare synthetic near-infrared spectra to the
observed GNIRS spectrum of \object, in an attempt to constrain
effective temperature \tf, surface gravity, and metallicity. If
\tf\ and log\,$g$ can be constrained then mass and age can also be
constrained, from evolutionary models.

The distance to \object\ is currently unknown and therefore, in
comparing against model spectra, the constraint provided by the
normalisation of the spectrum is not available. Together with the
limitations of the spectral models (principally opacity uncertainties,
described below) this precludes the accurate measurement of parameters
\tf, log\,$g$, and [m/H] for \object\ by direct comparison of
synthetic spectra to data alone. To overcome these limitations we have
chosen to use the recent detailed modelling of \mobject\ by
\citet{sau07} as a benchmark. Then to estimate parameters for \object,
we compare differences in the measured spectra between \mobject\ and
\object, with differences in the model spectra, as parameters are
changed away from the fiducial values for \mobject. Put another way,
the models are calibrated to \mobject. The rationale, and the method
itself, are similar to the procedure developed by \citet{bur06b}, but
with significant differences.

We first provide details of the spectral models, and then present the
results of the model fits. Combinations of \tf, log\,$g$, and [m/H]
are denoted e.g. [750, 5.0, 0.0].

\subsection{The BT-Settl models of T dwarf spectra}
\label{phoenix}

The model atmospheres used to generate the theoretical T dwarf spectra
for this comparison were generated with version 15.3 of the
general-purpose stellar atmosphere code \texttt{PHOENIX} \citep{jcam}.
For the present models we use a setup that gives the currently best
fits to observed spectra of M, L, and T dwarfs, updating the
microphysics used in the GAIA model grid
\cite[]{2005A&A...442..281K,2006A&A...452.1021K}.  

We adopt the solar
abundances of \citet{grev98}. Several recent studies of line formation
in the solar atmosphere have determined carbon, oxygen and nitrogen
abundances nearly a factor of two lower
\citep{asp04,agsSolAbun,apSolCO_CS14}.  However these revisions have
been found to be in conflict with other observations and models of the
solar atmosphere, such as studies of the carbon monoxide bands
\citep{ayresSolCO}, and in particular with helioseismology-based
models of the solar interior
\citep[e.\,g.][]{2006ApJ...649..529D}. The issue of the solar
abundance scale for C, O and probably N, is therefore far from
settled.  Since the opacity in an essentially dust-free T dwarf
atmosphere is largely dominated by the molecular bands of H$_2$O,
CH$_4$, and to some extent NH$_3$, it is the relative abundance of CNO
that primarily determines the structure and spectral energy
distribution of these objects.  A reduction of just these three
elements by a factor of two is thus almost equivalent to decreasing the
overall metallicity to [m/H]=$-$0.3 on the \citet{grev98} scale,
\ie\ the solar-metallicity models presented here are nearly identical
to metal-rich models of [m/H]=+0.3 based on the \citet{agsSolAbun}
solar abundances.

The equation of state (EOS) is an updated version of that used in
\citet{LimDust}, including $\sim 10^3$ species of atoms, ions and
molecules as well as allowing for the formation and dissolution of 
$\sim 100$ grain species. 
One major update improves on the `Cond' limit of
treating the EOS and the opacity of condensates, described in
\citet{LimDust}, which handled the absence of dust clouds from the
visible photosphere at low \tf\ by omitting their opacity
from the radiative transfer calculations, while still treating their
formation in thermodynamical and chemical equilibrium.  The more
realistic `Settl' treatment self-consistently includes the
gravitational settling, sedimentation or rainout, of condensates,
which results in a cloud layer retreating to larger optical depths as
\tf\ decreases \citep{all03,bdEkstasy,all07}.  In particular
this provides a more realistic description of the depletion of
refractory elements from the atmosphere.

The water line opacities are taken from the calculations of
\citet[BT]{bt2}, found to be giving the best overall fit to the water
bands over a wide temperature range. 

Obtaining accurate and complete opacity data of CH$_4$ and NH$_3$ has
generally been a much greater challenge. 
The current models include the latest linelists calculated
with the Spherical Top Data System \citep[STDS]{wc98} to account for
the principal infrared bands of methane \citep{hahbSTDS}. 
But coverage of the hot bands in the 
near-infrared, at 2.1--2.5\,$\mu$m and especially at
1.5--1.7\,$\mu$m, is limited since the higher vibrational levels
contributing to these transitions have yet to be modelled.
Semi-empirical corrections based on high-temperature simulations
and laboratory measurements \citep{bcjw02,2005JQSRT..96..251B} have
been included to account for this incompleteness. 
The situation is even worse for the 1.0--1.5\,$\mu$m region, where
only low-temperature linelists with mostly unknown ground state
energies are available and opacities at brown dwarf temperatures can
only be estimated.  NH$_3$ shows comparable problems starting already
at 1.9--2.1\,$\mu$m, with most of the lower states unknown for the
1.4--1.7\,$\mu$m bands. These models and the quality of the opacity
data are evaluated in more detail in comparison to the two late T
dwarfs \gobject\ and \mobject\ by \citet{2MASS04pap}.

Another important recent improvement in the \texttt{PHOENIX} models is
the inclusion of new atomic line profile data based on more accurate
interaction potentials. These profiles have improved the fits to
observed spectra significantly over the standard model of a Lorentzian
line shape at large detunings \cite[]{Alkalis03,BurrVolNaK}.  
We use a detailed and depth-dependent line profile for each of the
alkali resonance lines (Li, Na, K, Rb, Cs D1 and D2, respectively) in
our calculations.  
This is particularly important for the optical spectrum, 
being entirely dominated by line blanketing due to the
massively broadened Na\textsc{i} and K\textsc{i} doublets 
which in T dwarfs extend all the way out to the peak of the $Y$-band, 
\ie\ beyond 1\,$\mu${m}.  
In the models presented here we have used the alkali line profiles
described in \citet{Alkalis03,alkalisLi,Allard06,Cores07,Allard07a}
and \citet{Johnas07}, which give a much improved representation of the
details of these line shapes. 

The quality of the match of these so-called `BT-Settl' models to real
spectra may be gauged by examining Figs \ref{gnirsfull},
\ref{gnirszoom}, and \ref{datamodel}. In Fig.~\ref{datamodel} we
compare the best-fit model for \object\ determined from the spectral
comparison below. The fit is rather good. At somewhat higher
temperatures the inadequacy of the CH$_4$ opacities manifests itself
in the wavelength range $1.6-1.7\mu$m, as is particularly evident in
comparing the top and bottom panels in Fig.~\ref{gnirszoom}. Another
region where the fit is relatively poor is the wavelength range
$1.1-1.15\mu$m.

In their detailed modelling of three cool brown dwarfs \citet{sau06}
and \citet{sau07} have found evidence for non-equilibrium effects in
the carbon and nitrogen chemistry of their atmospheres due to the
dredge-up of CO and N$_2$ from deeper and hotter layers into the
photosphere by turbulent mixing. This is parameterized by an
eddy-diffusion coefficient $K_{zz}$ in their models.  From analysis of
the mid-infrared spectra they conclude that NH$_3$ is strongly diluted
in the upper atmosphere due to this effect, and CO is significantly
enhanced, requiring $K_{zz} \simeq 10^6$\,cm$^2$ \,s$^{-1}$. We have
included this quenching of the chemical equilibrium in a similar way,
deriving mixing timescales from the same convective overshoot velocity
field that is also used in our modelling of the cloud formation and
settling processes, and find a similar dilution of NH$_3$ in
the mid-infrared. The effects of this dilution are also evident in the
near-infrared bands of NH$_3$, leading to considerably less absorption
at 1.45\,--\,1.55\,$\mu$m and 1.9\,--\,2.1\,$\mu$m than otherwise
expected.

\begin{figure}
\includegraphics[width=8.5cm]{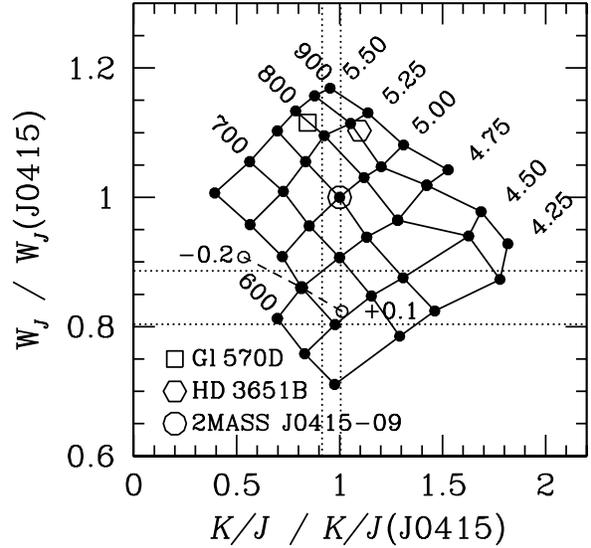}
\caption{Plot of the normalised index $W_J$, 
  quantifying the $J$-band peak width, 
  against the normalised index $K/J$, 
  measuring the ratio of the heights of the $K$- and $J$-band peaks. 
  The grid shows measured indices for 
  solar-metallicity spectral models, $600 \,{\rm K} \leq \tf \leq 900\,{\rm K}$,
  $4.25 \leq \log g \leq 5.50$,
  and is normalised by associating the model [750, 5.0, 0.0] 
  with \mobject, which is represented by the circle at (1.0, 1.0)
  (by definition). \gobject\ is represented by the square, and
  HD\,3651B by the hexagon.
  The horizontal and vertical bands plot the $\pm 2\sigma$ ranges of the
  normalised indices for \object. The effect of varying metallicity is
  illustrated by the dashed line.}
\label{tgplot}
\end{figure}

\subsection{Spectral comparison}
\label{speccomp}

\begin{table*}
\centering
\caption{Measured values of the spectral indices $W_J$ and {\em K/J}}
\begin{tabular}{@{}lcccc@{}}
\hline
 Index                    &  \gobject\ & \mobject\ & \object\ & HD\,3651B \\ 
                             &       &       &                 & \\ \hline
 $W_J$                       & 0.346 & 0.310 & 0.262$\pm$0.006 & 0.342\\
 $W_J$ / $W_J(J0415)$        & 1.116 & 1.000 & 0.845$\pm$0.021 & 1.103\\
{\em K/J}                    & 0.113 & 0.134 & 0.128$\pm$0.003 & 0.147\\
{\em K/J} / {\em K/J}(J0415) & 0.845 & 1.000 & 0.960$\pm$0.022 & 1.097
\\ \hline
\end{tabular}
\label{tab_newindices}
\end{table*}

To estimate the fundamental physical parameters for \object\ we use
the new model spectra, and follow a procedure similar to that
developed by \citet{bur06b}. These authors created a grid of
solar-metallicity spectral models, for which they measured a pair of
spectral indices: the H$_2$O$-J$ index (Table \ref{tab_indices}), and
an index {\em K/H} measuring the ratio of the heights of the {\em K}
and {\em H} peaks. The same indices were measured for \gobject\ for
which the properties \tf, log\,$g$, [M/H] are well determined
\citep{geb01,sau06}. The indices measured for the model with these
properties are not identical to the indices measured on the actual
spectrum, because the models are not perfect. For each of the two
indices, the ratio of the actual index over the model index provides a
scaling, which is used to calibrate the grid of model indices. The
properties \tf, log\,$g$, of any cool brown dwarf are then determined
by comparing the measured indices against the calibrated model
indices, and determining the best fit. An advantage of the method is
that it does not require the distance of the object to be known, since
the spectral indices record flux ratios. Nevertheless the effect on
the measured parameters of varying metallicity away from solar was not
explored (except for one source).

Interestingly, \citet{bur06b} estimated a value of $\tf \la 700$\,K for
the T8 dwarf 2MASS 0939$-$2448 and the T7.5 dwarf 1114$-$2618, lower
than the effective temperature determined for \mobject. Recently
\citet{leg07b} re-examined the spectra of these two sources in detail.
They argued that the best fits were achieved with $\tf \simeq 750$\,K, and
sub-solar metalicity [M/H]$\simeq-0.3$, and suggested that the method of
\citet{bur06b} needs to be revised to take metallicity into account.

In our comparison we also employ two indices. The first is the {\em
  K/J} index of \citet{bur06b}, defined as
\begin{equation}
K/J=\frac{\int F_{2.06-2.10}}{\int F_{1.25-1.29}}
\end{equation}
where $\int F_{\lambda_1-\lambda_2}$ denotes the integrated flux
between $\lambda_1$ and $\lambda_2$. For the second index we chose not
to use H$_2$O$-J$, because of the very limited dynamic range available
at temperatures \tf$<750$\,K, since the absorption is practically
saturated. Referring to Figs \ref{gnirsfull} and \ref{gnirszoom}, we
define an index $W_J$ which characterises the width of the {\em
  J}-band peak, and is defined by
\begin{equation}
W_J=\frac{\int F_{1.18-1.23}}{2\int F_{1.26-1.285}},
\end{equation}
where the factor of 2 in the denominator compensates for the wider band of
the numerator. 

The measured values of the two indices for the three sources \gobject,
\mobject, and \object\ are provided in Table \ref{tab_newindices}. For
interest we have also provided the measured values for the T7.5 dwarf
HD\,3651B \citep{mug06}. A brief discussion of this source is provided
at the end of this section, but for the remainder we
consider only the first three sources. The table includes the
measured indices, as well as the values normalised to the measured
indices of \mobject.  The published spectra of \gobject\ and
\mobject\ are of much higher S/N than the spectrum of \object\ we have
obtained, so the lack of error spectra for \gobject\ and \mobject\ is
unimportant in this analysis.  The {\em K/J} indices for the three
sources are all quite similar, while for $W_J$ the measured index
decreases through the sequence \gobject, \mobject, \object,
quantifying the progressive narrowing of the {\em J-}band peak visible
in Figs \ref{gnirsfull} and \ref{gnirszoom}.

The normalised indices from Table~\ref{tab_newindices} are plotted in
Fig.~\ref{tgplot}, where the dotted lines outline the $\pm2\sigma$
ranges for \object, along with a grid of solar-metallicity spectral
models.  The grid has been scaled to the data by associating the model
[750, 5.00, 0.0] with \mobject, and normalising the indices of all the
other models to the measured values for this model.  Choosing a
different model to represent \mobject\ merely has the effect of
shifting (and slightly distorting) the grid in this space such that
the new reference model lies at $(1.0, 1.0)$. In this way it can be
seen that the locations of \gobject\ and \object\ imply temperature
and gravity differences relative to \mobject\ that are, due to the
regularity of the grid, almost independent of the actual model chosen
to represent \mobject.  The plot illustrates the fact that the two
indices provide complementary information on the two parameters
\tf\ and log\,$g$, at fixed metallicity. The effect of decreasing
temperature alone is a decrease in both $W_J$ and $K/J$, while
decreasing gravity alone causes a decrease in $W_J$, but an increase
in $K/J$.

If \mobject\ and \object\ have similar metallicity, the plot indicates
that \object\ has \tf\ between 60 and 120\,K cooler, and log $g$
between 0.25 and 0.5 lower. Because only two indices are used, the
three parameters \tf, log\,$g$, [M/H] cannot be determined
independently. The effect of varying metallicity is illustrated in
Fig.~\ref{tgplot}, by reference to the model [650, 4.75, 0.0]. The
dashed line through this model shows the effect of varying the
metallicity over the range $-0.2<\Delta$[m/H]$<+0.1$. Because the line
approximately follows the line of constant gravity, varying
metallicity has little effect on the temperature estimate, but,
rather, is nearly degenerate with gravity. This degeneracy may be
described by the relation $\Delta$(log\,$g)\equiv-2\Delta$[m/H]. Therefore
\object\ is of lower gravity or higher metallicity than \mobject\ as
described by the limits
$-0.5<\Delta$(log\,$g-2$[m/H]$)<-0.25$.\footnote{\citet{liulc}
  find a similar amplitude for the degeneracy between
  metallicity and surface gravity determination, based on empirical
  comparison of the T7.5 dwarfs Gl 570D and HD\,3651B and on
  examination of the multi-metallicity models of \citet{burrows06}.}

The properties of \mobject\ were determined by \citet{sau07} to lie
along a narrow line from {\tf, log\,$g$} of {725, 5.00} to {775,
  5.37}. Adopting {750, 5.20} for \mobject\ implies $630<\tf<690$
($\pm2\sigma$ range) for \object. The temperature estimate from the
spectral modelling is in good agreement with the estimate derived from
the {\em H}$-$[4.49] colour of $560<\tf<660$\,K ($\pm2\sigma$
range). Combining these two independent estimates gives an estimate
$\tf=648\pm13$ ($1\sigma$ error), where the uncertainty reflects the
random errors in the GNIRS spectrum, and the mid-infrared photometry,
but takes no account of the uncertainties in the spectroscopic models,
in the extrapolation of the temperature-colour relation
(Fig. \ref{tempcol}), and in the temperature of \mobject. On the other
hand since the spectroscopic models are calibrated to \mobject\ at
750\,K, it seems very unlikely that \object\ is hotter than 700\,K. We
adopt the conservative ranges $600<\tf<700$\,K, and
$4.5<$log\,$g<5.1$, assuming solar metallicity, as the final estimate
of the effective temperature and gravity of \object. Confirmation of
this low effective temperature will require mid-infrared spectroscopy,
and measurement of the parallax, both of which are scheduled.

One of the models within the above ranges is [650, 4.5, 0.0]. The
synthetic spectrum generated with these parameters reproduces the
observations quite well (given the known opacity problems), as shown
in Fig.~\ref{datamodel}.

Finally we note that the indices listed in Table \ref{tab_newindices}
for \gobject\ and for HD\,3651B suggest that the latter is some 50\,K warmer.
\citet{liulc} argued that HD 3651B is of higher gravity
than \gobject, in which case the relative values of the two indices
indicate that HD 3651B must also have higher metallicity, in agreement
with their findings.

\begin{figure*}
\includegraphics[width=17cm]{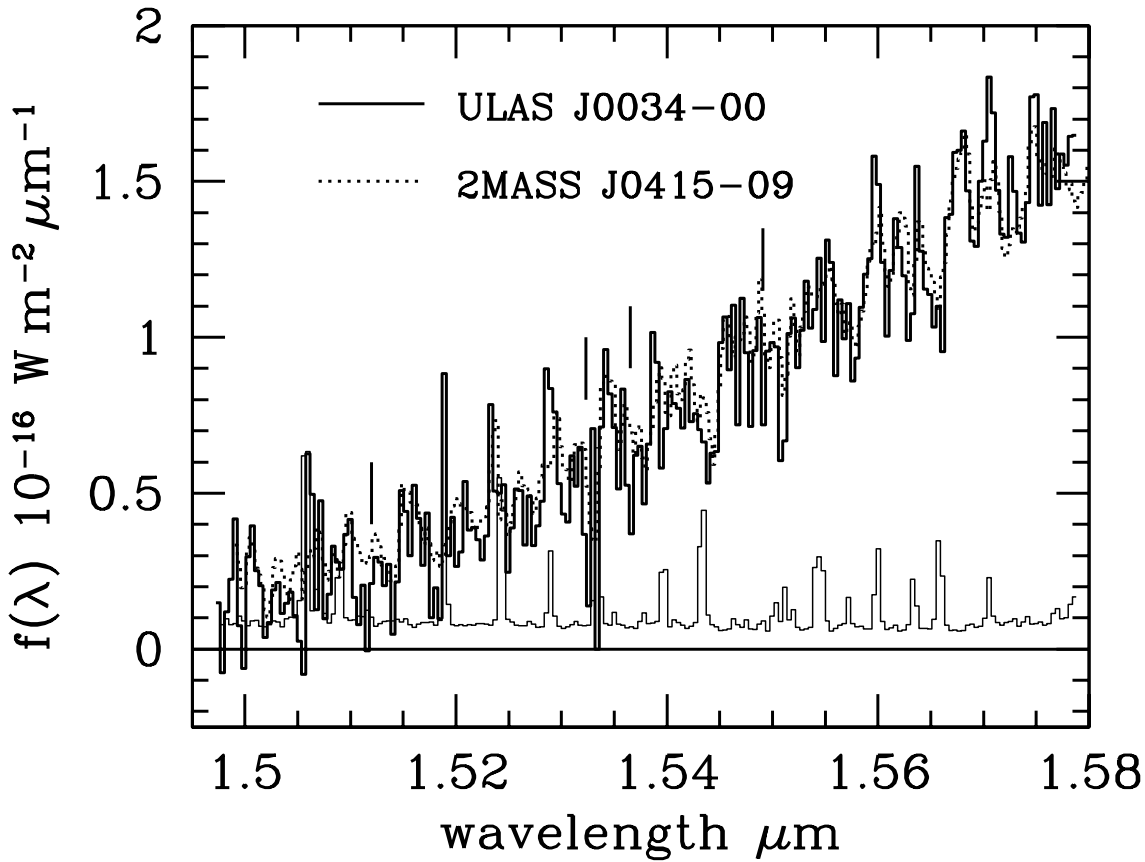}
\caption{Comparison of the $H-$band spectra of \object\ and \mobject,
  at intermediate resolution. The thick solid line is the ISAAC
  spectrum of \object\, binned by a factor of five, to a pixel size
  4.0\AA. The thin solid line is the corresponding error spectrum. The
  dotted line is the spectrum of \mobject\ from \citet{mclean}.}
\label{isaac}
\end{figure*}

\subsection{Distance, age and mass}

\citet[their Fig.~3]{sau07} plot isochrones and isomass sequences as
a function of \tf\ and log\,$g$.  They also demonstrate that
photospheric metallicity does not significantly impact the
evolutionary parameters in this diagram. The ranges in \tf\
and log\,$g$ of 600--700\,K and 4.5--5.1, determined above,
correspond to ranges in [age (Gyr), mass ($M_{\rm Jup}$)] of [0.5, 15] to
[8, 36]. Hence \object\ is likely to be less massive and less luminous
than 2MASS J0415$-$09, for which Saumon et al. determine a mass of 
33--58\,$M_{\rm Jup}$.
 
The distance of \object\ may be estimated from the modelled surface
fluxes and brown dwarf radii.  Normalising the models to the peak of
the $J$ band implies a distance of 14--24\,pc for the ranges in
effective temperature and surface gravity given above.  This is
consistent with the observed correlation between absolute magnitude
and spectral type shown for example in Figs. 8 and 9 of
\citet{kna04}. The measured proper motion of
$(0\farcs37\pm0\farcs07)$/yr then translates into a tangential
velocity of 25--42\,km\,s$^{-1}$, suggesting membership of the younger
disk population \citep[e.g.][]{egg98}, which is consistent with the
age range of 0.5--8\,Gyr derived from the temperature and gravity
above.

\section{Search for NH$_3$ in the $H$-band}
\label{ammonia}

Absorption by NH$_3$ is potentially a valuable diagnostic of the
conditions in the atmospheres of very cool brown dwarfs
\citep{burrows03}. NH$_3$ absorption was first convincingly detected
in a mid-infrared spectrum of $\epsilon$ Indi Bab \citep{roellig}, and
subsequently in all the dwarfs of spectral type T2 or later, observed
by \citet{cush06} in the mid-infrared. The search for and the
interpretation of NH$_3$ absorption in the near-infrared is made
difficult by the shortcomings in the line lists referred to
previously. \citet{sau00} report the detection of very weak NH$_3$
absorption in the {\em H} and {\em K} bands in the spectrum of the T7p
dwarf Gl229B. The weakness of the features was interpreted as a
consequence of nonequilibrium chemistry. This finding is corroborated
by analysis of NH$_3$ absorption in the mid-infrared spectrum of
\gobject\ \citep{sau06}. At lower temperatures NH$_3$ should
strengthen, and its detection would be a valuable aid in further
developing the spectral models of dwarfs with \tf$<700$\,K. The NH$_3$
absorption may develop into a sufficiently strong spectral feature
\citep{burrows03} that it will be incorporated into future schemes for
classification beyond T8, possibly motivating the creation of a new
spectral type (but see discussion in \citet{leg07b}).  Our motivation
for obtaining a higher-resolution spectrum in the {\em H} band was to
investigate whether the excess absorption seen in the low-resolution
spectrum in the blue wing of the {\em H-}band peak is resolved into
absorption lines of NH$_3$ at higher resolution.

The ISAAC spectrum of \object\ is reproduced in Fig.~\ref{isaac}. For
the purposes of the plot the spectrum has been binned by a factor of
five, to reduce the noise, so the plotted spectrum is undersampled
relative to critical sampling by a factor of about two. The thin line
shows the error spectrum for this binning.  When further binned to
match the sampling of the GNIRS spectrum, the two spectra are in close
agreement, confirming the excess absorption relative to \mobject. The
dotted line in the figure plots the intermediate resolution spectrum
of \mobject\ from \citet{mclean}, scaled to match \object\ over the
wavelength range $1.56-1.575\mu$m. The overall match between the two
spectra is rather close, in terms of detailed features. Nevertheless,
a general depression of the spectrum of \object\ shortward of
$1.55\mu$m is evident. To assess the reality of any possible
absorption lines we subtracted the spectrum of \mobject. We then ran a
boxcar filter of width seven (original) pixels (5.6\AA) over the
difference spectrum and the variance spectrum, and created a S/N
spectrum for an absorption line of this width. Three candidate
absorption lines at S/N$>4$ were detected, at wavelengths
1.5323\,$\mu$m, 1.5365\,$\mu$m (the most significant), and
1.5491\,$\mu$m.  These three lines are marked in Fig.~\ref{isaac}.
Using a broad 50\,\AA\ filter, we also detect a significant feature
centred near 1.512 $\mu$m. This lies near the wavelength of a feature
in the NH$_3$ laboratory spectrum \citep{leg07b}, although there are
also many strong lines of H$_2$O absorption in this region. While
these lines appear to be real, and could be due to NH$_3$, they are
quite weak. There is no obvious strong feature apparent in the
spectrum which would suggest that \object\ is radically different,
requiring a new spectral type. \citet{leg07b} have recently addressed
the question of where the strongest NH$_3$ features may appear in the
spectra of very cool brown dwarfs, and concluded that the {\em Y}-band
and {\em J}-band peaks may be a more promising region to search.

\section{Summary}
\label{summ}

We have reported the discovery of the T dwarf \object, discovered in
UKIDSS DR1, which has a spectral type of T8.5 and an effective
temperature of $600<\tf<700$\,K, making it the latest and coolest T
dwarf known at the present time.  In particular it is $\sim 100$\,K
cooler than \mobject\ and probably also has lower luminosity and
lower mass.  Despite the low inferred temperature, an
intermediate-resolution spectrum in the $H$-band failed to reveal
convincingly the presence of absorption by NH$_3$, although similar
spectroscopy of the $Y$- and $J$-band peaks may be more revealing.  It
will be extremely valuable to measure the parallax of \object, in
order to confirm the estimated temperature without recourse to
somewhat uncertain spectroscopic models. High-resolution imaging
of \object\ will be similarly important to establish whether it is
isolated.

The area covered by the UKIDSS LAS will expand by a factor of 20 over
the next few years, so prospects for finding even cooler brown dwarfs
are extremely good.  However the unexpectedly blue $Y-J$ colour of
\object\ should be taken into account in designing future searches for
such ultra-cool dwarfs.  While all the current models predict that
ultra-cool T dwarfs become bluer in \jh\ with decreasing temperature
\citep{he06}, they are somewhat discrepant in their predictions for
how \yj\ varies with temperature: the models of \citet{burrows03}
initially tend towards redder \yj; those of \citet{marley02} show a
weak trend to bluer \yj; and the BT-Settl models show a stronger trend
to bluer \yj\ down to $\sim 550$\,K.  Given the relative depths in the
{\yjhk} bands in UKIDSS \citep{wa07a}, searches for very cool brown
dwarfs should allow for the possibility that they are detected only in
the $Y$ and $J$ bands \---\ one of several issues that should be
clarified as more of these objects are discovered.

%%%%%%%%%%%%%%%%%%%%%%%%%%%%%%%%%%%%%%%%%%%%%%%%%%%%%%%%%%%%%%%%%%%%%%%%%%%%%%

\section*{Acknowledgments}

We are grateful to the anonymous referee for a very detailed report,
including numerous suggestions that have improved the presentation.
We would also like to thank the staffs of the Joint Astronomy Centre,
Hawai'i, of the Cambridge Astronomical Survey Unit, and of the
Astronomy Technology Centre and the Wide Field Astronomy Unit in
Edinburgh, all of whom contributed to the creation of UKIDSS. We thank
the ESO Director General and the Director of the {\em Spitzer Space
  Telescope} for awards of Director's Discretionary Time. Results
reported here are based on observations obtained at the Gemini
Observatory, through program GS-2006B-Q-36. Gemini Observatory is
operated by the Association of Universities for Research in Astronomy
Inc. (AURA), under a cooperative agreement with the NSF on behalf of
the Gemini partnership: the National Science Foundation (United
States), the Particle Physics and Astronomy Research Council (United
Kingdom), the National Research Council (Canada), CONICYT (Chile), the
Australian Research Council (Australia), CNPq (Brazil) and CONICET
(Argentina).  SKL is supported by the Gemini Observatory, operated by
AURA on behalf of the international Gemini partnership. MCL
acknowledges support for this work from NSF grants AST-0407441 and
AST-0507833 and an Alfred P. Sloan Research Fellowship.

%%%%%%%%%%%%%%%%%%%%%%%%%%%%%%%%%%%%%%%%%%%%%%%%%%%%%%%%%%%%%%%%%%%%%%%%%%%%%%

\bsp

\label{lastpage}

\end{document}